\documentstyle[12pt]{article}

\let\ssection=\section
\def\section{\setcounter{equation}{0}\ssection}

\font\frbig=eufm10  scaled\magstep1
\font\frscr=eufm7 scaled\magstep1
\font\frscrscr=eufm5 scaled\magstep1
\newfam\frfam
\textfont\frfam=\frbig
\scriptfont\frfam=\frscr
\scriptscriptfont\frfam=\frscrscr
\def\fr{\fam\frfam}

\font\openbig=msym10  scaled\magstep1
\font\openscr=msym7 scaled\magstep1
\font\openscrscr=msym5 scaled\magstep1
\newfam\openfam
\textfont\openfam=\openbig
\scriptfont\openfam=\openscr
\scriptscriptfont\openfam=\openscrscr
\def\open{\fam\openfam}

\font\Scbig=cmss10  scaled\magstep1
\font\Scscr=cmss8 scaled\magstep1
\font\Scscrscr=cmss8
\newfam\Scfam
\textfont\Scfam=\Scbig
\scriptfont\Scfam=\Scscr
\scriptscriptfont\Scfam=\Scscrscr
\def\Sc{\fam\Scfam}

\makeatletter

\newdimen\normalarrayskip
\newdimen\minarrayskip
\normalarrayskip\baselineskip
\minarrayskip\jot
\newif\ifold\oldtrue\def\new{\oldfalse}
\def\arraymode{\ifold\relax\else\displaystyle\fi} 

\def\@arrayskip{\ifold\baselineskip\z@\lineskip\z@
\else
\baselineskip\minarrayskip\lineskip2\minarrayskip\fi}
\def\@arrayclassz{\ifcase\@lastchclass\@acolampacol\or
\@ampacol\or\or\or\@addamp\or
\@acolampacol\or\@firstampfalse\@acol\fi
\edef\@preamble{\@preamble
\ifcase\@chnum
\hfil$\relax\arraymode\@sharp$\hfil
\or $\relax\arraymode\@sharp$\hfil
\or\hfil$\relax\arraymode\@sharp$\fi}}
\def\@array[#1]#2{\setbox\@arstrutbox=\hbox{\vrule
height\arraystretch\ht\strutbox
depth\arraystretch\dp\strutbox
width\z@}\@mkpream{#2}\edef\@preamble{\halign\noexpand\@halignto
\bgroup\tabskip\z@\@arstrut\@preamble\tabskip\z@\cr}%
\let\@startpbox\@@startpbox\let\@endpbox\@@endpbox
\if #1t\vtop\else\if#1b\vbox\else\vcenter\fi\fi
\bgroup\let\par\relax
\let\@sharp##\let\protect\relax
\@arrayskip\@preamble}
\makeatother

\def\lvm{\leavevmode\hbox to\parindent{\hfill}}
\def\tilde{\widetilde}
\def\bar{\overline}
\def\a{\alpha}
\def\o{\omega}
\def\j{\psi}
\def\ff{\varphi}

\def\req#1{(\ref{#1})}
\def\dd#1{{\partial\over\partial{#1}}}
\def\L{\left}
\def\R{\right}
\def\Rarrow{\rightarrow}

\def\BE{\begin{equation}}
\def\EE{\end{equation}}
\def\BA{\begin{array}}
\def\EA{\end{array}}

\def\hat{\widehat}
\def\half{{1\over2}}
\def\res{{\rm res}}
\def\tr{{\rm tr}}
\def\d{\partial}
\def\ddsc#1{{\partial^2\over\partial{#1}^2}}
\def\pr{^\prime}

\def\emt{energy-momentum tensor}
\def\cft{conformal field theory}
\def\h{hierarchy}
\def\hs{hierarchies}
\def\cs{constraints}
\def\K{Kontsevich}
\def\M{Miwa}
\def\mm{matrix model}
\def\mms{matrix models}
\def\V{Virasoro}
\def\RS{Riemann surface}

\begin{document}
\hfuzz=1pt

\date{{\small corrected version\\April 10, 1992}}
\title{\sc Solving Virasoro Constraints on Integrable Hierarchies via
the \K-\M\ Transform}

\author{{\large A.~M.~Semikhatov}\\
{\small {\sl Theory Division, P.~N.~Lebedev Physics Institute}}\\
{\small {\sl 53 Leninsky prosp., Moscow 117924, Russia}}}

\maketitle

\begin{abstract}
We solve \V\ constraints on the KP \h\ in terms of minimal conformal
models. The constraints we start with are implemented by the \V\
generators depending on a background charge $Q$. Then the solutions to the
constraints are given by the theory which has the same field content as
the David-Distler-Kawai theory: it consists of a minimal matter scalar
with background charge $Q$, dressed with an extra `Liouville' scalar. In
particular, the \V-constrained tau function is related to the correlator
of a product of (dressed) `21' operators. The construction is based on a
generalization of the \K\ parametrization of the KP times achieved by
introducing into it \M\ parameters which depend on the value of $Q$. Under
the thus defined \K-\M\ transformation, the \V\ \cs\ are proven to be
equivalent to a master equation depending on the parameter $Q$. The master
equation is further identified with a null-vector decoupling equation. We
conjecture that $W^{(n)}$ constraints on the KP hierarchy are similarly
related to a level-$n$ decoupling equation. We also consider the master
equation for the $N$-reduced KP hierarchies. Several comments are made on
a possible relation of the generalized master equation to {\it scaled}
Kontsevich-type matrix integrals and on the form the equation takes in
higher genera.\end{abstract}

\newpage
\section{Introduction and Discussion}\lvm The
claim of the Matrix Models approach \cite{[BK],[DSh],[GM]} is a
non-perturbative description of two-dimensional gravity (and
gravity-coupled matter). The main computational tool is provided in
applications \cite{[DVV],[GGPZj],[DfK]} by the theory of integrable
\hs\ subjected to the \V\ constraints
\cite{[D],[FKN1],[DVV],[MM],[IM]}. The relevance of the \V-constrained
\hs\ to the intersection theory on the moduli space of curves has
also been proven in \cite{[W1],[K],[W2]}. The \V\ \cs\ thus
constitute a fundamental notion of the theory and are the heart of
\mms' applications to both gravity-coupled theories and the intersection
theory\footnote{The case studied in most detail is the \V-constrained
KdV \h\ whose relation to the intersection theory on moduli space
of Riemann surfaces has been discussed in \cite{[W1]}.}. A major task is
to find their general solution (see, for instance, \cite{[Sc]}).

On the other hand, a challenging problem remains of giving a direct
proof of the equivalence between the `hierarchical' formalism and the
\cft\ description of quantum gravity \cite{[KPZ],[DK],[Da]}. As a
`direct proof' one would like to have something more than just the
circumstantial evidence. It may seem discouraging that assuming an
equivalence between the conformal-theory description of quantum gravity
and the theory of (appropriately constrained) integrable hierarchies,
one then has to believe that certain ingredients of \cft\ satisfy
integrable equations, while these seem to be a long way from the
equations which are known to hold for \cft\ correlators
\cite{[BPZ],[DF],[KZ]}.

We will show in this paper that \V\ \cs\ on the KP hierarchy
are {\it solved} by minimal models, by virtue of the equations
\cite{[BPZ],[DF]} satisfied by the corelators.

As has been understood for quite some time, the only viable candidate
for a `space-time' for the conformal theory underlying integrable
\hs\ to live on, could be the spectral curve associated to the
hierarchy. Yet the attempts to actually build up such a theory were
hindered by a problem that remained: the infinite collection of time
parameters inherent to integrable \hs\ are hard to deal with
within the standard conformal techniques.

Fortunately, there does exist a formalism in which the time variables
are treated, in a sense, on equal footing with the spectral parameter.
This is the \M\ transform used in the KP \h\ \cite{[Mi],[Sa]}.
What is more, a similar construction has been introduced by \K\
\cite{[K]} in his \mm \footnote{The \K\ \mm\ is important by itself --
it provides a combinatorial model of the universal moduli space
\cite{[K]} and, as such, serves as an important step in demonstrating
the KdV \h\ in the intersection theory on the moduli space -- and
it also provides a model of quantum gravity \cite{[W2]}. It is not,
however, of the form of the \mms\ considered previously, which raises
the question of its equivalence to one of the ``standard" models. The
crucial point in studying this equivalence is, again, the proof of the
\V\ \cs\ satisfied by the \K\ matrix integral \cite{[W2],[MMM],[IZ]}.
Once the \cs\ are established, one is left with ``only" the proof
that they specify the model uniquely.}. (More recently, it has been used
in \cite{[W2],[MMM],[GN]} in relation to the \V\ \cs\ on integrable
hierarchies, although in a context different from the one considered in
this paper.)

It turns out that in order to relate the \V\ \cs\ on the KP
\h\ to certain conformal field theory data, one needs to introduce
additional parameters into the \K\ parametrization of the KP times. That
is, the \K\ parametrization can be viewed as a special case of \M's,
and the extra `degrees of freedom' present in the \M\ transform should
not be completely frozen: by varying them one moves between {\it different}
(generalised) \K\ transformations. We will see that different \K\
transformations should be used depending on the operators under
consideration. One thus gets a `\M-parametrized set' of \K\
transformations, which is referred to below as the \K-\M\ transform.

For each of the generalised \K\ transformations, pulling back the \V\
\cs\ results in relations, analogous to the ``master equation''
of ref.\cite{[MS]} (see also ref.\cite{[MMM]}), which are satisfied
\cite{[S34]} by correlation functions of an `auxiliary' \cft\ provided
it contains a null-vector \cite{[BPZ]}. This \cft\ therefore gives a
solution to the \V-constrained hierarchy.
It consists of a `matter' scalar $\varphi$ with the \emt $$
T_{\rm m}=-\half \d\varphi\d\varphi+{i\over 2}Q_\d^2\varphi, $$
{\it and} an {\it extra} scalar $\phi$. The background charge $Q$ enters
in the \V\ generators on the KP hierarchy. In order for the \K-\M\
transform of the \V\ \cs\ to exist, the background charge must be related
to the \M\ parameter $n_i$ via
$$ Q={1\over n_i}-2n_i,$$
which allows us to interprete $n_i$ as half the cosmological constant.
When this condition is satisfied, the \V\ \cs\ map under the \K-\M\
transform into the decoupling equation \cite{[BPZ]} for the level-2 null
vector. The application of the classical technique of \cite{[BPZ]} to
\V-constrained \hs\ thus allows us to relate the \V-constrained KP \h\ to
the formalism of refs.\cite{[DK],[Da]}.

The relation which we find,
between \V\ \cs\ and null-vector decoupling equations is very instructive
from the point of view of string field theory. The decoupling equations
acquire the r\^ole of the sought string field theory equations, and
therefore at least for the matter central charge $d<1$, ``conformal
models provide classical solutions to the string field theory'' inasmuch
as the corresponding correlators satisfy the decoupling equations.

Another implication of the identification of the \V\ \cs\ with
the decoupling equation has to do with recursion relations in
topological theories. Recall that the recursion relations are
essentially the \V\ (or higher $W$-) constraints. Therefore the
particular form, obtained below, of the decoupling equations serves as a
generating function for the recursion relations. Reversing the
argument, it is amusing to know that certain \cft\ correlators are in
fact \V-constrained tau functions.

Also deserves mentioning the relation between the master equation and
higher-genus \RS s. We will show that the master equation, considered on
a coordinate patch of a \RS, can be extended to the whole of the \RS\ as
an equation for a certain `constituent' of the tau function.

There are several important
issues which should be further elucidated. The first one is the relation of
\V-constrained tau functions to \K-type matrix integrals. The cases
considered in the literature \cite{[K],[MS],[KMMMZ]} seem to apply only
when $\a^2=2$, while we would like to have matrix integrals that give
rise to more general master equations. Moreover, there is an evidence
that such matrix integrals pertain to the Toda lattice
hierarchy, which is a `discrete' hierarchy, and that the master
equations we are considering in this paper, follow only after a certain
scaling limit. This scaling limit can be viewed as an adaptation for the
spectral parameter of the scaling \cite{[S29]} of the \V-constrained
\hs\ Toda$\Rarrow$KP. Second, a possible relevance of higher \V\
null vectors to the \V-constrained KP \h\ points to a relation
between \V\ null vectors and $W$ algebras, the fact observed in a
different approach \cite{[BDfIZ]}. We conjecture that $W^{(n)}$ \cs\ on
the KP hierarchy give rise to a level-$n$ decoupling equation.

Further, the KP \h\ can be reduced to higher generalized `KdV'
\hs\ \cite{[DDKM],[DS]}. In this paper we will consider only the
series associated to the $sl(N)$ Kac-Moody algebras (which correspond to
the $A$-series minimal models \cite{[CIZ]}), and we will call these the
$N$-KdV hierarchies\footnote{\V\ \cs\ on the $N$-KdV hierarchies
admit a unified treatment, which is in turn a specialization of a
general construction applicable to \hs\ of the $r$-matrix type
\cite{[S33]}.}. Although neither the interpretation of \V-constrained
$N$-KdV \hs\ in terms of moduli spaces, nor the corresponding
\K-type matrix integrals are known, we will show that the `master
equation' can be naturally extended to this case as well.

We thus begin in Sect.~2 with fixing our notations and recalling some
basic facts about the \V\ action on the KP hierarchy. In Sect.~3 we
introduce the \K-\M\ transform and use it in order to recast the \V\
\cs\ into the ``master equation''. The inverse transform, from
the master equation to the \V\ \cs, is also proven here.
Further, to give the master equation
a conformal field-theoretic interpretation, we recall in Sect.~4 the
necessary elementary formulae pertaining to the decopupling of null
vectors in conformal models. After that, we establish in Sect.~5 a
relation between \V-constrained tau functions evaluated at different
values of the Miwa parameters, and the conformal field correlators. In
Sect.~6 we suggest a version of the master equation for the $N$-reduced
(generalized KdV) hierarchies. Sect.~7 presents a preliminary discussion
of the relation of our formalism to matrix integrals and the scaling in
the \K\ parametrization, as well as a possible r\^ole of higher null
vectors \footnote{It has been shown recently
\cite{[GS]} that the level-3 decoupling give rise to $W^{(3)}$
constraints on the tau function.}.
We also show how the master equation can be given
meaning on higher-genus \RS s.

\section{\V\ action on the KP hierarchy} \subsection{}\lvm
The KP \h\ is described in terms of $\psi {\rm Diff}$ operators
\cite{[DDKM]} as an infinite set of mutually commuting evolution
equations\BE {\d K\over\d t_r}=-(KD^rK^{-1})_-K,\quad r\geq1
\label{KPeqs}\EE on the coefficients $w_n(x,~t_1,~t_2,~t_3,\ldots)$ of
a $\psi{\rm Diff}$ operator $K$ of the form (with $D=\d /\d x $)\BE K
=1+\sum_{n\geq 1}w_n D^{-n}\label{(1)}\EE The wave function and the
adjoint wave function are defined by\BE\psi (t,z)=Ke^{\xi (t,z)}
,\quad\psi^\ast(t,z)=K^{\ast-1}e^{-\xi (t,z)},\quad\xi (t,z)
=\sum_{r\geq 1}t_r z^r\label{(2)}\EE where $K^\ast$ is the formal
adjoint of $K$. The wave functions are related to the tau function via
\BE\psi(t,z)=e^{\xi (t,z)}{\tau (t-[z^{-1}])\over\tau (t)},\quad
\psi^\ast(t,z)=e^{-\xi (t,z)}{\tau (t+[z^{-1}])\over\tau (t)}
\label{(3)}\EE where $t\pm [z^{-1}]=(t_1\pm z^{-1}, t_2\pm\half
z^{-2}, t_3\pm {1\over 3} z^{-3},\ldots)$.

\subsection{}\lvm Now we introduce a \V\ action on the tau function
$\tau(t)$: The \V\ generators read, \BE\new
\BA{rcl} {\Sc L}_{p>0} &=&\half\sum^{p-1}_{k=1}{\d^{2}\over\d
t_{p-k}\d t_k}+\sum_{k\geq 1}kt_k {\d\over\d t_{p+k}}+ (a_0+ (J-\half
)p) {\d\over\d t_p}\\ {\Sc L}_0&=&\sum_{k\geq 1}kt_k {\d\over\d t_k}+
{1\over 2}a^2_0-\half\L(J-\half\R)^2\\ {\Sc L}_{p<0} &=&\sum_{k\geq
1}(k-p)t_{k-p}{\d\over\d t_k}+\half
\sum^{-p-1}_{k=1}k(-p-k)t_kt_{-p-k}+ (a_0+ (J -\half )p)(-p)t_{-p}\EA
\label{Lontau}\EE They satisfy the algebra\BE [{\Sc L}_p, {\Sc L}_q ]
=(p-q) {\Sc L}_{p+q}-\delta_{p+q,0}(p^3-p)(J^2-J+ {1\over 6})
\label{(2.10)}\EE which shows, in particular, the role played by the
parameter {\it J}. Introducing the `\emt'\BE {\Sc T}(u)
=\sum_{p\in{\open Z}}u^{-p-2}{\Sc L}_p\label{ScT(u)}\EE we can deform
the tau function as\BE\tau(t)\mapsto\tau(t)+\delta\tau(t)=\tau (t)+{\Sc
T}(u)\tau (t)\label{(2.12)}\EE This action can be translated into the
space of dressing operators $K$. The result is \cite{[S10]} that $K$
gets deformed by means of a left multiplication,\BE\delta K=-{\fr
T}(u)K,\EE where ${\fr T}(u)$ is the \emt\ in the guise of a
pseudodifferential operator\footnote{We have chosen the irrelevant
parameter $a_0=J-\half$, see \cite{[S10]}.}
\BE{\fr T}(u)=(1-J){\d\psi(t,u)\over\d u}\circ D^{-1}\circ
\psi^\ast(t,u)-J\psi(t,u)\circ D^{-1}\circ{\d\psi^\ast(t,u)\over\d u}
\label{frT(u)}\EE Thus, ${\fr T}(u)$ reproduces the structure of the
\emt\ of a spin-$J$ $bc$ theory $(1-J)\d b\cdot c-Jb\cdot\d c$
\cite{[FMS]}. Expanding ${\fr T}(u)$ in powers of the variable $u$,
which was introduced in \req{ScT(u)} and has now acquired the role of a
spectral parameter, as\BE{\fr T}(u)=\sum_{p\in {\open Z}}u^{-p-2}{\fr
L}_p\label{(2.14)}\EE we arrive at the individual \V\ generators (which are a
particular case of the general construction applicable to integrable
\hs\ of the $r$-matrix type \cite{[S33]})\BE{\fr L}_n\equiv
K(J(n+1)D^n+ PD^{n+1})K^{-1})_-,\quad P\equiv x+\sum_{r\geq 1}rt_r
 D^{r-1}\label{frLn}\EE These define the \V-constrained KP hierarchy
via ${\fr L}_{n\geq -1}=0$.

Conversely, the \V\ generators \req{Lontau} acting on the tau function
can be recovered starting from the ${\fr L}_n$, eq.\req{frLn}, by using
the equation\BE\res K=-\d\log\tau,\label{resK}\EE
whence\BE\delta\d\log\tau=-\res\delta K=\res{\fr T}(u)K=\res{\fr
T}(u)\label{resKnext}\EE The (operator) residue of ${\fr T}(u)$ is
immediately read off from \req{frT(u)}. To the combination of the wave
functions thus appearing we apply the well-known formula\BE {\tau
(t-[u^{-1}]+ [z^{-1}])\over\tau (t)}=(u-z) e^{\xi (t,z)-\xi (t,u)}
\d^{-1}\L(\psi (t,u)\psi^\ast(t,z)\R)\label{tauovertau}\EE
The generators \req{Lontau} now follow by expanding
this at $u\Rarrow z$,

\section{Kontsevich--Miwa transform}\lvm The \M\ reparametrization of
the KP times is accomplished by the substitution\BE t_r={1\over
r}\sum_j n_jz^{-r}_j,\quad r\geq1\label{Miwatransform}\EE where
$\{z_j\}$ is a set of points on the complex plane and the parameters
$n_j$ are integer classically; we will need, however, to continue off
the integer values.

This parametrization puts, in a sense, the times and the spectral
parameter on equal ground. By the {\it\K} transform we will understand
the dependence, via eq.\req{Miwatransform}, of $t_r$ on the $z_j$ for
{\it fixed} $n_j$. Note that the way \K\ has used a parametrization of
this type implied setting all the $n_j$ equal to a constant which was 1
in ref.\cite{[MMM]}. In our approach this will prove too strong a
restriction. We thus proceed with the general $n_j$ and then find how
the $n_j$ must be tuned.

Generally, the \K-\M\ parametrization turns out very inconvenient with
regard to the use of the standard machinery of the KP hierarchy
(instead, the \M\ parametrization has been used to construct a quite
different, ``discrete'' formalism for the KP and related hierarchies
\cite{[Mi],[Sa]}). This applies also to the above \V\ generators. That
is, viewing \req{Miwatransform} as \BE t_r={1\over r}\int_{{\open
CP}^1}\!d\mu(z)n(z)z^{-r}\EE one could define the wave functions formally
as \BE\psi[n](z)=\prod_j\L(1-{z\over z_j}\R)^{\!-n_j}{1\over\tau [n]}
e^{-{\delta\over \delta n(z)}}\tau[n]\label{(3.4)}\EE However, using
this in the \emt\ \req{frT(u)} and similar formulae would require making
sense out of expressions such as ${\d\over\d z}{\delta\over\delta
n[z]}$. Even this would not be quite satisfactory, though, as one would
still have had to express the result in terms of the derivatives with
respect to the \K\ parameters $z_j$\/: for us, the tau function must be
a function $\tau\{z_j\}$ of points scattered over ${\open CP}^1$. The
$\d/\d z_j$ derivatives, however, are not easy to get hold of using the
equation \req{Miwatransform}.

There are two circumstances that save the day. First, we are interested
not in all the \V\ generators, but rather in those with non-negative
(and, in addition, $-1$) mode numbers ${\fr L}_{n\geq-1}$ (which are
used to define the \V-constrained \h\ via ${\fr L}_n=0$, $n\geq
-1$). Picking these out amounts to retaining in ${\fr T}(z)$ only terms
with $z$ to negative powers, $i.e.$, the terms vanishing at
$z\Rarrow\infty$. This part of ${\fr T}(z)$ is singled out as \BE {\fr
T}^{(\infty)}(v)=\sum_{n\geq-1}v^{-n-2}{1\over 2\pi i}\oint dz z^{n+1}
{\fr T}(z)={1\over 2\pi i}\oint dz{1\over v-z}{\fr T}(z)
\label{Tinfty}\EE where $v$ is from a neighbourhood of the infinity and
the integration contour encompasses this neighbourhood.

Second, a crucial simplification will be achieved by evaluating ${\fr
T}^{(\infty)}(v)$ only at the points from the above set $\{z_j\}$ (one
has to take care that they be inside the chosen neighbourhood). We thus
have to evaluate the operator ${\cal T}(z_i)$ from (see
eqs.\req{resK},\req{resKnext}) \BE\d\L({\cal T}(z_i)\tau\R)={1\over 2\pi
i}\oint dz{1\over z_i-z}\res{\fr T}(z)\label{defcalT}\EE This will
depend on the collection of the $n_j$, which we will indicate by a
subscript. It is straightforward to find that \BE\new\BA{rcl} {\cal
T}_{\{n\}}(z_i) &=& {1\over 2\pi i} \oint dz { 1\over z_i-z}\L\{(J-\half
){1\over z}\sum_{r\geq 1}z^{-r-1}{\d\over\d t_r}+ \half
\sum_{r,s}z^{-r-s-2} {\d^2\over\d t_r\d t_s}\R.\\
{}&+&\L.\sum_j{n_j\over z_j-z}\sum_{r\geq 1}z^{-r-1}{\d\over\d t_r}+
\half \sum_j{n_j+n_j^2\over (z_j-z)^2}+ \half \sum_{^{j,k}_{j\neq
k}}{n_jn_k\over (z_j -z)(z_k -z)}\R.\\ {}&-&\L. J\sum_j{n_j\over
(z_j-z)^2}+ (J-\half )\sum_{r\geq 1}z^{-r-2} r{\d\over\d t_r}\R\}
\EA\label{(3.8)}\EE where we have substituted \req{Miwatransform} for
each explicit occurence of the $t_r$.

However, the problem is that we need all the $\d/\d t_r$-derivatives to
be expressed in terms of the $\d/\d z_j$ as well, while the equation
relating $t_r$ and $z_j$ does not allow this. It is only when we
evaluate the contour integral in \req{(3.8)} that the $t$-derivatives
will arrange into the combinations which are just the desired
$\d/\d{z_j}$'s. As the integration contour encompasses all the points
$\{z_j\}$, the residues at both $z=z_i$ and $z=z_j$\/, $j\neq i$,
contribute to \req{(3.8)}. The residue at $z_i$ consists of the
following parts: first, the terms with simple poles contribute
\BE \new
\BA{l}\L(J-\half -{1\over 2n_i}\R) {1\over n_i}{1\over z_i}{\d\over\d
z_i}-{1\over 2n_i^2}{\d^2\over\d z_i^2}+{1\over n_i}\sum_{j\neq i}
{n_j\over z_j-z_i}{\d\over\d z_i}\\ -\L( J-\half -{1\over
2n_i}\R)\sum_{r\geq 1}r z^{-r-2}_i {\d \over\d t_r}-\half \sum_{j\neq
i}{n_j+n _j^2-2Jn_j\over (z_j-z_i)^2} -\half \sum_{{j\neq i\atop {k\neq
i\atop k\neq j}} }{n_jn_k\over (z_j - z_i)(z_k-z_i)} \EA\label{(3.9)}
\EE where we have substituted \BE \new \BA{rcl}
\sum_{r,s\geq1}z^{-r-s-2}_i {\d^2\over\d t_r\d t_s} &=& {1\over n_i^2
}{\d^2\over\d z_i^2 }+ {1\over n_i^2}{1\over z_i} {\d\over\d
z_i}-{1\over n_i}\sum_{r\geq 1}z_i^{-r-2} r {\d\over\d t_r},\\
\sum_{r\geq 1}z^{-r-1}_i {\d\over \d t_r} &=&-{1\over n_i}{\d\over\d
z_i}\EA\label{derivatives}\EE

In the third term inside the curly brackets, a second-order pole occurs
when $j=i$, which produces
\BE{1\over z_i}\dd{z_i}-n_i\sum_{r\geq 1}rz_i^{-r-2}\dd{t_r}\EE
Next, second-order poles occur in the double sum over $j$, $k$ in
\req{(3.8)}: \BE {1\over 2\pi i}\oint dz { 1\over z_i-z} \sum_{j\neq
i}{n_j n_i\over (z_j-z)(z_i-z)}=\sum_{j\neq i}{n_i n_j\over (z_i-z_j)^2}
\EE

We thus see that the term $\sum_{r\geq1}rz_i^{-r-2}\dd{t_r}$, which cannot
be expressed locally through $\dd{z_j}$, enters with the coefficient
$-\L(J-\half-{1\over 2n_i}+n_i\R)$. We have to set this coefficient to
zero; therefore $n_i$ and $J$ are related by
\BE{1\over n_i}-2n_i=2J-1\equiv Q\label{ni}\EE
Then the contribution of the residue at $z=z_i$ becomes \BE \new
\BA{rcl} {\cal T}_{\{n\}}^{(i)}(z_i)=&-&{1\over 2n_i^2 }{\d^2\over\d
z_i^2 }+ {1\over n_i}\sum_{j\neq i}{n_j\over z_j-z_i} {\d\over\d z_i}\\
{}&-& \half \sum_{{j\neq i\atop {k\neq i\atop k\neq j}} }{n_jn_k\over
(z_j-z_i)(z_k-z_i)} - \half \sum_{j\neq i}{n_j+n_j^2-2Jn_j -2n_i
n_j\over (z_j-z_i)^2} \EA\label{(3.12)}\EE Similarly, each of the
residues at $z_j$, $j\neq i$, contributes \BE {{\cal
T}_{\{n\}}}_{(j)}(z_i)=-{1\over z_j-z_i}{\d\over\d z_j}+ {1\over
z_j-z_i}\sum_{k\neq j}{n_j n_k\over z_k-z_j} +\half
{n_j+n_j^2-2Jn_j\over (z_i-z_j)^2}\label{(3.13)}\EE and thus, finally
\footnote{We have used the identity $$\sum_{j\neq
i}\sum_{{k\neq i}\atop {k\neq j}}{1\over
(z_j-z_i)}{n_jn_k\over(z_k-z_j)}=\half \sum_{j\neq i}\sum_{{k\neq
i}\atop {k\neq j}}{n_jn_k\over (z_j-z_i)(z_k-z_i)}$$},
\BE \new \BA{rcl} {\cal T}_{\{n\}}(z_i) &=& {\cal
T}_{\{n\}}^{(i)}(z_i)+\sum_{j\neq i}{{\cal T}_{\{n\}}}_{(j)}(z_i)\\
{}&=& -{1\over 2n_i^2 }{\d^2\over\d z_i^2 }+ {1\over n_i}\sum_{j\neq
i}{1\over z_j-z_i}\L( n_j {\d\over \d z_i}-n_i {\d\over\d z_j}\R)
\EA\label{Tgen}\EE
where $n_i$ is to be determined from \req{ni}.

If one wishes {\it all} the ${\fr T}^{(\infty)}(z_j)$ to carry over to
the \K\ variables along with ${\fr T}^{(\infty)}(z_i)$, all the $n_j$
have to be fixed to the same value $n_i$. Then, one gets
``symmetric'' operators \BE
{\cal T}(z_i)=-{Q^2+4\pm Q\sqrt{Q^2+8}\over 4}
{\d^2\over\d z_i^2 }-\sum_{j\neq i}{1\over
z_j-z_i}\L({\d\over\d z_j}-{\d\over\d z_i}\R)\label{Tsym}\EE
These differential operators, of course, satisfy the centreless algebra
spanned by the $\{n\ge-1\}$-\V\ generators.

Clearly, if one starts with the {\it \V-constrained} KP hierarchy, {\it
i.e.}, ${\fr T}^{(\infty)}(z)=0$, one ends up in the \K\ parametrization
with the KP \V\ {\sl master equation} (cf. ref.\cite{[MS]}) ${\cal
T}(z_i).\tau\{z_j\}=0$. In the next section we show that this is solved
by certain \cft\ correlators.

Conversely, let us also see how, given the master equation ${\cal
T}_{\{n\}}(z_i).\tau\{z_j\}=0$, one can recover the usual form of the \V\
\cs. The required transformation is inverse to the one we have just
performed, and its less trivial part is to get rid of the explicit
occurences of the $z_j$. The derivatives $\d/\d z_j$, on the other hand,
are straightforwardly replaced with $\d/\d t_r$ according to
\req{derivatives}. We thus get\BE{\cal T}_{\{n\}}(z_i)\!=\!-{1\over
2n_i^2}\L(\!n_i^2\!\sum_{r,s\geq 1}z_i^{-r-s-2}{\d^2\over\d t_r\d
t_s}+n_i\!\sum_{r\geq 1}z_i^{-r-2} (r+1)\dd{t_r}\!\R)+\sum_{j\neq
i}\!\sum_{r\geq 1}n_j{z_j^{-r-1}-z_i^{-r-1}\over z_j - z_i}
\dd{t_r}\label{*}\EE In the last term, we divide by $z_j-z_i$ and get
$-\sum_{j\neq i}n_j\sum_{r\geq
1}\sum_{s=1}^{r+1}z_j^{-s}z_i^{-r-s+2}\dd{t_r}$. Now the sum over {\it
all \/} $j$ gives $st_s$ according to the \M\ transform
\req{Miwatransform}; the missing term with $j=i$, which is to be added and
subtracted, combines with the term of the same structure from \req{*}. We
thus recover the ($\geq-1$) \V\ generators \req{Lontau} for
$J$ given by \req{ni}.

\section{Trivialities on \cft}\lvm Now, to prepare the presentation in
the next section, consider the subject which is apparently quite
different from what we have had so far. Introduce a conformal theory of
a $U(1)$ current and an \emt: \BE j(z)=\sum_{n\in{\open Z}}j_n
z^{-n-1},\quad T(z)=\sum_{n\in {\open Z}}L_n z^{-n-2}\label{jT}\EE
\BE\new\BA{lrcl} \L[ j_m ,\R.&\L. j_n\R] &=& km\delta_{m+n,0}\\\L[ L_m
,\R.&\L.\!\!\!\!L_n\R] &=& (m-n)L_{m+n}+ {d+1\over 12}(m^3 -
m)\delta_{m+n,0}\\ \L[ L_m ,\R.&\L.\!\!\!\!j_n\R] &=& -nj_{m+n}
\EA\label{thetheory}\EE
(We have parametrized the central charge as $d+1$). Let $\Psi$ be a
primary field with conformal dimension $\Delta$ and $U(1)$ charge $q$.
Then, by a slight variation of \cite{[BPZ]}\footnote{We extend the standard
system of \cite{[BPZ]} by introducing a current, but then `compensate' the
extra freedom by suppressing a term (the one proportional to
$j_{-1}^2$) in the general form of the decoupling equation.},
we find that the level-2
state \BE |\Upsilon\rangle=\L(\alpha L_{-1}^2+ L_{-2}+\beta
j_{-2}+\gamma j_{-1} L_{-1}\R)|\Psi\rangle\label{Upsilon}\EE
is primary provided \BE \alpha={k\over 2q^2}~,\quad\beta=-{q\over
k}-{1\over 2q}~,\quad\gamma=-{1\over q}~,\quad\Delta=-{q^2\over
k}-\half\label{parameters}\EE with $q$ given by, \BE{q^2\over
k}={d-13\pm\sqrt{(25-d)(1-d)}\over 24}\label{q2}\EE and,
accordingly,\BE\Delta={1-d\mp\sqrt{(25-d)(1-d)}\over 24}~.\label{Delta}
\EE

Factoring out the state $|\Upsilon\rangle$ leads in the usual manner to
the equation \BE \L\{ {k\over 2q^2}{\d^2\over\d z^2 }-{1\over
q}\sum_{j} {1\over z_j-z}\L(q{\d\over\d z_j} - q_j{\d\over\d
z}\R)+{1\over q}\sum_{j}{q\Delta_j-q_j \Delta\over (z_j-z)^2}\R\}
\langle\Psi (z)\Psi_1 (z_1)\ldots\Psi_n (z_n)\rangle=0
\label{decouplinggen}\EE
where $\Psi_j$ are primaries of dimension $\Delta_j$ and $U(1)$ charge
$q_j$. In particular, \BE \L\{ {k\over 2q^2}{\d^2\over\d z_i^2
}+\sum_{j\neq i} {1\over z_i-z_j}\L({\d\over\d z_j}-{\d\over\d
z_i}\R)\R\}\langle\Psi (z_1)\ldots\Psi (z_n)\rangle=0
\label{decoupligsym}\EE These equations will be crucial for comparison
with the KP \h\ in Sect.~5.

Writing the Hilbert space as (matter$)\otimes($current$)\equiv {\cal
M}\otimes {\cal C}$, $|\Psi\rangle=|\psi\rangle\otimes |\tilde
{\Psi}\rangle$, we introduce the matter \V\ generators $l_n$ by, \BE
L_n=l_n+\tilde{L}_n\equiv l_n+ {1\over 2k}\sum_{m\in {\open
Z}}:j_{n-m}j_m :\label{(23)}\EE
They then have central charge $d$. Now, using in \req{Upsilon} that \BE
\tilde{L}_{-1}|\tilde {\Psi}\rangle={q\over k}j_{-1}|\tilde
{\Psi}\rangle , \quad \tilde{L}_{-2}|\tilde {\Psi}\rangle=\L({q\over
k}j_{-2}+{1\over 2k}j_{-1}^2\R)|\tilde {\Psi}\rangle ,\quad
\tilde{L}_{-1}^2|\tilde {\Psi}\rangle=\L({q\over k}j_{-2}+{q^2\over
k^2}j_{-1}^2\R)|\tilde {\Psi}\rangle\EE we find \BE |\Upsilon\rangle
=\L({k\over 2q^2}l_{-1}^2+ l_{-2}\R) |\Psi\rangle\label{(24)}\EE
Therefore we are left with a null vector in the matter Hilbert space
${\cal M}$. The dimension of $|\psi\rangle$ in the matter sector is
found from \BE L_0 |\Psi\rangle=\L(l_0+ {1\over 2k}j_0^2\R)
|\Psi\rangle\EE and equals \BE
\delta=\Delta-{1\over 2k}q^2={5-d\mp\sqrt{(1-d)(25-d)}\over 16}
\label{delta}\EE
which for the appropriate values of $d$ is of course the dimension of
the `21' operator of the minimal model with central charge $d$.

\section{\it A la r\'echerche de Liouville perdu} \subsection{Ansatz for
the \V-constrained tau function}\lvm
A contact between sections 4 and 3, i.e., between \cft\ formalism and
the KP \h\ is suggested by the above derivation of the `master'
operators \req{Tgen},\req{Tsym}, in which the $z_j$
were viewed as coordinates on the spectral curve.

For the \V-constrained tau function in the \K\ parametrization we
assume the ansatz \BE\tau\{z_j\}=\lim_{n\Rarrow\infty}\langle\Psi
(z_1) \ldots\Psi (z_n)\rangle\label{ansatz}\EE Then, comparing the
decoupling equation \req{decoupligsym} with the master equation
\req{Tsym}, we find
\BE n_i^2={q^2\over-k}\EE
and therefore, taking into account \req{ni} and \req{q2},
\BE Q=\sqrt{{1-d\over3}}\equiv Q_{\rm m}\EE
The \M\ parameter $n_i$ is determined in terms of the central charge $d$
as (with $\sigma$ being a conventional sign factor, $\sigma^2=1$)
\BE n_i=\sigma{\mp Q+\sqrt{Q^2+8}\over4}\equiv-\sigma{-Q_{\rm
L}\pm Q_{\rm m}\over 4}=-{\sigma\over2}\a_{\pm}\EE
where $Q_{\rm L}$ and $Q_{\rm m}$ are recognized as the Liouville and the
matter central charge respectively, and $\a$ is the cosmological constant.

Note that the
\emt\ $T(z)$ from \req{jT} appears to have a priori nothing to do with
${\fr T}(z)$ (or, which is the same, with ${\Sc T}^{(\infty)}(z_i)$, see
\req{ScT(u)}) we have started with. In terms of the latter tensor,
the master operator \req{Tgen} comprises contributions of all the
positive-moded \V\ generators, while out of $T(z)$ only $L_{-1}$ and
$L_{-2}$ are used in the construction of $|\Upsilon\rangle$,
eq.\req{Upsilon}.

\subsection{More general correlators}\lvm
To reconstruct matter theory field operators, consider
the form the ${\Sc L}_{n\geq -1}$-\V\ \cs\ take for the wave
function of the hierarchy, $w(t,z_k)\equiv e^{-\xi (t,z_k)}\psi(t,z_k)$,
which should now become a function of the $z_j$, $w\{z_j\}(z_k)$. More
precisely, consider the `unnormalized' wave function
$\bar{w}\{z_j\}(z_k)=\tau\{z_j\}w\{z_j\}(z_k) $. Then the use of the
\K\ transform at the \M\ point $n_j=\a/2$, $j\neq k$ and $n_k=-1$,
gives\footnote{To obtain the insertion into the correlation function
\req{barw} at the point $z_k$ of the operator we are interested in by
itself, rather than its fusion with the `background' $\Psi$, we use the
\K\ transform at the value of the \M\ parameter $n_k=-1$ instead of
$n_i-1$. This means that we are in fact considering
$\bar{w}\{z_j\}^{}_{j\neq k}(z_k)$. Similar remarks apply to other
correlation functions considered below. Of course, the conceptual
difference between the tau function and the `unnormalized' wave
functions ${\bar w}(z_k)$ disappears in the \M\ parametrization.}
\BE\bar{w}\{z_j\}(z_k)=\L\langle\prod_{j\neq k}
\Psi(z_j)\cdot\Xi(z_k)\R\rangle\label{barw}\EE
where $\Xi$ is a
primary field with the $U(1)$ charge $q/n_i$ and dimension
$\Delta/n_i=-\sigma{2\Delta\over\a}$.
This implies in turn that its dimension in the
matter sector equals
\BE-\sigma{2\Delta\over\a}-{1\over2k}\L({q\over n_i}\R)^2
=(\mp)\half Q+\half\equiv\L\{\BA{l} 1-J\\J\EA\R.\EE
Thus the wave
function is related to, say, (depending on the sign conventions)
the $b$-field of the $bc$ system \footnote{For the values of $J$ that we
will actually need (which are not half-integer nor even rational), the
$bc$ system would be purely formal. We will not keep it in the `$bc$' form
for long and bosonize it shortly.}. The
adjoint wave function is then similarly related to the corresponding $c$
field: for instance, the function $\tau(t-[z^{-1}_k]+[z^{-1}_l])$ is
annihilated by the operator \BE -{2\over\a^2}{\d^2\over\d z_i^2 } -
\sum_{{j\neq i,~j\neq k}\atop{j\neq l}}{1\over z_j-z_i} \L({\d\over\d
z_j}-{\d\over\d z_i}\R)+{2\over\a}\L({1\over z_l-z_i}-{1\over
z_k-z_i}\R){\d\over\d z_i}\EE
Again, we interpret this as a decoupling
equation which accounts for the effect of certain insertions at $z_k$ and
$z_l$. We thus find that the tau function $\tau (t-[z^{-1}_k]+
[z^{-1}_l])$ is proportional to the correlation function
\begin{eqnarray}\lefteqn{\L\langle \prod_{{j\neq k}\atop{j\neq
l}}\Psi(z_j)\exp{\L({q\over kn_i}\int^{z_k}\!j\R)}b(z_k)\exp{\L(-{q\over
kn_i}\int^{z_l}\!j\R)}c(z_l)\R\rangle}\nonumber&&\\
&&\qquad\qquad{}=\L\langle\prod_{{j\neq k}\atop{j\neq l}} \Psi(z_j)
(z_k-z_l)\exp{\L({q\over kn_i}\int^{z_k}_{z_l}\!j\R)}B(z_k)C(z_l)\R\rangle
\label{BC} \end{eqnarray} where we have used $$-{q^2\over
k^2n_i^2}j(z)j(w)\sim-{q^2\over k^2n_i^2}k\ln (z-w)=\ln(z-w).$$
Note that, although it is tempting to
take in \req{BC} the limit $z_k \Rarrow z_l$, this cannot be done
naively, as it would affect the whole construction of the \K-\M\
transform~!

As a cross-check, it is interesting to compare \req{BC} with the
identity \req{tauovertau} (which is valid for a general (i.e., not
necessarily \V-constrained) KP tau-function). In the \K\
parametrization we have \BE e^{\xi (t,z)-\xi
(t,u)}=\prod_j\L({z_j-u\over z_j-z}\R)^{n_j}\EE On the other hand,
fusing the exponential in \req{BC} with the product of the $\Psi(z_j)$,
gives the factor \BE\prod_{{j\neq k}\atop{j\neq l}}\L({z_j-z_k\over
z_j-z_l}\R)^{{q\over k}\L(-{q\over kn_i}\R)\cdot k}=\prod_{{j\neq
k}\atop{j\neq l}}\L({z_j-z_k\over z_j-z_l}\R)^{n_i}\EE which agrees
with the above now that the $n_j$ have been set in \req{ansatz} equal to
$n_i$. This suggests extending the
\K-\M\ transform `off-shell', i.e., off the \V\ constraints. Both of the
two classes of objects, the tau function etc., and the theory in ${\cal
M}\otimes{\cal C}$, exist by themselves, while we have seen that imposing
the \V\ \cs\ on the one end results in factoring over a submodule on the
other.

By bosonizing the formal $bc$ system one gets a matter scalar $\varphi$
with the familiar \emt\ (\cite{[BPZ],[DF],[FQS]}) \BE T_{\rm
m}=-\half \d\varphi\d\varphi+{i\over 2}Q_{\rm m}\d^2\varphi .
\label{Tmatter}\EE Note that for the unitary series \BE d=1-{6\over
p(p+1)},\EE the \M\ parameter $n_i$ is determined as \BE n_i^2=\half
\L({p+1\over p}\R)^{\mp1}.\EE

Further, as to the theory in ${\cal C}$, recall that we have
\BE [j_m,j_n]=km\delta_{m+n,0},\quad j_{n>0}|\Psi\rangle=0,\quad
j_0|\Psi\rangle=q|\Psi\rangle\EE with negative $q^2/k$\ \footnote{for
$d<1$. For $d>25$, on the other hand, $q^2/k$ is positive, but then one
has to consider the \h\ with imaginary $n_i$~! It appears that the
matter and the Liouville theory then take place of one another, and
$n_i=\sqrt{-1}\a/2$ with $\alpha$ being the cosmological constant.}. To
see what the current corresponds to in the KP theory, consider the
correlation function with an extra insertion of an operator which
depends on only $j$: \BE\L\langle\prod_{j\neq k\atop j\neq l}\Psi(z_j)
\exp\L((\pm){Q\over \sqrt{-k}}\int^{z_k}_{z_l}j\R)\R\rangle
\label{expj}\EE The decoupling equation states that this is annihilated
by the operator \BE {\cal T}(z_i)+\half Q(Q\pm Q_{\rm L})\L({1\over
z_k-z_i}- {1\over z_l-z_i}\R) {\d\over\d z_i}\EE and therefore coincides,
up to a constant, with the \V-constrained tau function $\tau (t)$
evaluated at the \M\ point \BE n_j=\L\{\BA{ll}-{\sigma\over 2}\a, &j\neq
k,\quad j\neq l\\ Q,&j=k\\-Q,&j=l\EA\R.\EE
i.e., this is
$$\tau\biggl(-{\sigma\over2}\a\sum_{{j\neq k}\atop{j\neq
l}}[z^{-1}_j]+Q_[z^{-1}_k] -Q_[z^{-1}_l]\biggr)$$ This is
another illustration of how the \K-\M\ transform works: establishing the
relation to different conformal field operators ${\cal O}(z_k)$ requires
fixing different values of $n_k$.

\subsection{The dressing prescription}\lvm The balance of
dimensions and $U(1)$ charges of both the $\Psi$ and $\Xi$ operators
follows a particular general pattern. That is, as there are no
$1/(z_i-z_j)^2$-terms in the master equation, we have to ensure that these
terms be absent from the decoupling equation \req{decouplinggen}.
Therefore we can only consider
operators from a special sector, i.e., those whose dimensions $\Delta_j$
and $U(1)$ charges $q_j$ satisfy (see \req{decouplinggen})\footnote{we
continue to denote by $\Delta$ and $q$ {\it the} dimension and $U(1)$
charge from \req{thetheory} -- \req{Delta}, i.e., those of $\Psi$.},
\BE
\Delta_j=\Delta{q_j\over q}=(\pm)\half Qn_j
\label{Deltaj}\EE (Clearly, the $U(1)$ charges are related to the \M\
parameters via $q_j/\sqrt{-k}=n_j$.) With this condition, the decoupling
equation \req{decouplinggen} takes the form,
\BE\L\{-{2\over\a^2}
{\ddsc{z_i}}+\sum_{j\neq i}{1\over z_j-z_i}\L(-{2\sigma\over\a}n_j{\d\over
\d z_i} - {\d\over\d z_j}\R)\R\}\L\langle\Psi(z_i)\prod_{j\neq
i}\Psi_j(z_j)\R\rangle=0\label{decouplingn}\EE and we are still able, as
before, to relate this to the \V\ constraints, since the operator on the
LHS is precisely the general form of the `master' operator \req{Tgen} in
which only one value, that of $n_i$, out of the $n_j$, has been fixed.
Let us repeat once again that evaluating the tau function at different
\M\ parameters $\{n_j\}$ corresponds to different operator insertions in
the \cft\ language.

Now, the dimension of the matter part of $\Psi_j$ is equal to
\BE\delta_j=\Delta_j -{q_j^2\over 2k}=\Delta{q_j\over q}-{q_j^2\over
2k}\label{deltaj}\EE On the other hand, dimensions of the matter field
operators $e^{i\gamma\ff}$ are fixed by the \emt\ \req{Tmatter}. It is
crucial for consistency that the two formulae agree: as the term linear
in $q_j/\sqrt{-k}$ enters in \req{deltaj} with the coefficient
(sign)$\half Q$, eq.\req{deltaj} will always be satisfied for the
matter operators $e^{i\gamma\varphi}$ provided
$q_j/\sqrt{-k}=(\pm)\cases{\gamma\cr Q-\gamma\cr}$.
Therefore, the prescription for the `dressing'
inherited from the \V-constrained KP \h\ says that the
coefficients in front of the two scalars $\varphi$ and $\phi$ in the
exponents coincide up to the reflection $\gamma\mapsto Q_{\rm m}-\gamma$
in the matter sector (and, to be precise, up to the usual overall factor
of $i$). Thus, although the field content is the same as in
ref.\cite{[DK]}, it is not quite the David-Distler-Kawai formalism that
follows directly from the KP hierarchy\footnote{This can be seen also by
noticing that the dimensions in ${\cal M}$ and ${\cal C}$ do not add up
to 1; nor is the central charge equal to 26. This is not a surprise,
since the current $j$ is not anomalous.}. Our `dressing exponent' \BE
\frac{q_j}{\sqrt{-k}}=-\frac{\sqrt{-k}\Delta}{q}\pm\frac{1}
{2\sqrt3}\sqrt{1-d+24\delta_j}\EE
differs from eq.(3.12) of \cite{[DK]} \BE \beta_j=-\frac{1}{2}Q_{\rm
L}\pm\frac{1} {2\sqrt3}\sqrt{1-d+24\delta_j}\EE by the cosmological
constant $\a$.

Equivalently, the `bulk' dimensions $\Delta_j$, rather than being equal
to 1, are related to the gravitational scaling dimensions of fields.
Indeed, evaluating the gravitational scaling dimension of $\psi$
according to \cite{[Da],[DK],[KPZ]}, \BE\hat{\delta}_{\pm}=
{\pm\sqrt{1-d+24\delta}-\sqrt{1-d}\over\sqrt{25-d}-\sqrt{1-d}}\EE one
would find \BE \hat{\delta}_+={3\over 8}\pm
{d-4-\sqrt{(1-d)(25-d)}\over 24}\label{hat delta}\EE with the sign on
the RHS corresponding to that in \req{q2} and the subsequent formulae.
In particular, choosing the {\it lower} signs throughout, we have
$\hat{\delta}_+=\Delta+ \half $. More generally, the gravitational
scaling dimensions corresponding to \req{deltaj} equal \BE \hat
{\delta}_{j+}=-{q_j q\over k}=\Delta_j+\half{q_j\over
q}=\Delta_j
-{\sigma\over\a}{q_j\over\sqrt{-k}}\EE and thus are given by the
$\Delta_j$ `corrected' by the terms linear in the charge.

The combination $-n_k\sum_{r\geq1}z_k^{-r-1}\d/\d t_r=\d/\d z_k$ of the
$\d/\d t_r$-derivatives applied to the decoupling equation gives rise to
the recursion relations for correlators of
$\sum_{r\geq1}z_k^{-r-1}\sigma_r$. Interestingly, it is therefore the
decoupling equation that serves as a generating relation for the
recursion relations.

\section{The $N$-reduced equations}\lvm In this section we show how the
master equation can be obtained for the $N$-reduced case. The KP
\h\ can be reduced to generalized $N$-KdV \hs\
\cite{[DDKM]} by imposing the constraint \BE Q^N\equiv L\in{\rm Diff}
\label{QN}\EE requiring that the $N^{\rm th}$ power of the Lax operator
be purely differential. Then, in a standard manner, the evolutions along
the times $t_{Nk}$\/, $k\geq 1$, drop out and these times may be set to
zero. The rest of the $t_n$ are conveniently relabelled as $t_{a,i}=
t_{Ni+a}$, $i\geq 0$, $a=1,\ldots,N-1.$

As to the \V\ generators, only ${\fr L}_{Nj}$ out of the generators
\req{frLn} are compatible with the reduction in the sense that they
remain symmetries of the reduced \h\ without imposing further
\cs\ \cite{[S29]}. The value of $J$ can be set to zero
\cite{[S29]}, and thus we arrive at the generators \BE{\fr
L}^{[N]}_j={1\over N}\L(K\L(x+\sum_{a,i}(Ni+ a)t_{a,i}D^{N(i+j)+a}\R)
K^{-1}\R)_{\!-}\label{(2.17)}\EE which span a \V\ algebra of their
own. To construct the corresponding \emt, recall that the spectral
parameter of the $N$-KdV \h\ is $\zeta=z^N$. Then \BE \new
\BA{rcl} {\fr T}^{[N]}(\zeta)(d\zeta)^2 &\equiv&\sum_{j\in{\open
 Z}}\zeta^{-j-2}{\fr L}^{[N]}_j(d\zeta)^2\\ {}&=& N\L(
K\sum_{b,j}(Nj+b) t_{b,j} D^{Nj+b}{1\over z^2}\delta (D^N,z^N)
K^{-1}\R)_{\!-}\!(dz)^2\EA\label{(2.18)}\EE Now, $\delta(z,D)$ is a
projector onto an eigenspace of $D$ with the eigenvalue $z$, and thus
\BE \delta (D^N,z^N)={1\over N}\sum^{N-1}_{c=0}\delta (z^{(c)}, D),\quad
z^{(c)}=\omega^cz,\quad \omega=\exp\L({2\pi\sqrt{-}1\over N}\R)
\label{deltaN}\EE Using this we bring the above \emt\ to the form \BE
{\fr T}^{[N]}=\sum^{N-1}_{c=0}\omega^c{\d\psi(t,z^{(c)})\over\d
z}\circ D^{-1}\circ\psi^\ast(t,z^{(c)})={1\over
N}\sum^{N-1}_{c=0}\omega^{2c}{\fr T}(z^{(c)})\label{frT[N]}\EE where we
have used that the spectral parameter of an $N$-KdV \h\ lies on a
complex curve defined in ${\open C}^2\!\ni\! (z,E)$ by an equation
$z^N\!=\!P(E)$. Then, $\psi$ and $\psi^\ast$ are defined on this curve,
and after the projection onto ${\open CP}^1$ yield $N$ wave functions
$\psi^{(a)}(t,E)$, distinct away from the branch points. That is, we
have defined \BE \psi^{(a)}(t,E)=Ke^{\xi (t,z^{(a)})}\equiv
w(t,z^{(a)})e^{\xi (t,z^{(a)})},\qquad\xi (t, z^{(a)})
=\sum_{j,b}t_{b,j} (z^{(a)})^{Nj+b}\EE Note a similarity between
\req{frT[N]} and the \emt\ of conformal theories on ${\open Z}_N$-curves
\cite{[BR]}.

The \V\ action on the tau-function of the $N$-reduced \h\ can be
recovered using the eqs.\req{resK} -- \req{tauovertau}, combined with
taking the appropriate average over ${\open Z}_N$. In this way we arrive
at the \V\ generators \BE \new\BA{rcl} n > 0: \qquad {\Sc L}^{[N]}_n
&=& {1\over N}{1\over 2}\sum^{N-1}_{a=1}\sum^{n-1}_{i=0}{\d^2\over\d
t_{a,i}\d t_{N-a,n-i-1}}+ {1\over N}\sum^{N-1}_{a=1}\sum_{i\geq 0}(Ni+
a)t_{a,i}{\d\over\d t_{a,i+n}},\\ {\Sc L}^{[N]}_0 &=& {1\over
N}\sum^{N-1}_{a=1}\sum_{i\geq 0}(Ni+ a)t_{a,i} {\d\over\d t_{a,i}},\\ n
 < 0: \qquad {\Sc L}^{[N]}_n &=& {1\over N}{1\over 2}\sum^{N-1
}_{a=1}\sum^{-n-1}_{i=0}(Ni+a)(-N(i+n)-a) t_{a,i} t_{N-a,-i-n-1}\\
{}&+&{1\over N}\sum^{N-1}_{a=1}\sum_{i\geq -n} (Ni+a)t_{a,i}{\d\over\d
t_{a,i+n}} \EA\label{(ScL[N])}\EE

Now, we have learnt from the above derivation of the matter operator
\req{Tgen} that $z_i$ is nothing but a value taken by the spectral
parameter. Therefore the trick with averaging over ${\open Z}_N$ as in
\req{deltaN} can be carried over to the \K\ parametrization. That is, to
perform the reduction to an $N$-KdV hierarchy, it suffices to substitute
\BE z_i \mapsto \omega^cz_i
\EE and then sum over ${\open Z}_N$ as in \req{frT[N]}. Indeed, having
defined the reduced ${\cal T}$-operator as \BE\d\L({\cal
T}^{[N]}_i\tau\R)={1\over 2\pi i}\oint dz {1\over z_i-z}{1\over N}\sum
^{N-1}_{c=0}\omega^{2c}\res{\fr T}(\omega^cz),\EE one continues this as
\BE\new\BA{rcl}={1\over 2\pi i}{1\over N}\sum^{N-1}_{c=0}\oint
{dz\omega^{-
c}\over z_i-\omega^{-c}z}\omega^{2c}\res{\fr T}(z)&=&{1\over
2\pi i}{1\over N}\sum^{N-1}_{c=0}\omega^{2c}\oint{dz\over\omega^cz_i -
z}\res{\fr T}(z)\\ {}&=&\d\L({1\over N}\sum^{N-1}_{c=0}\omega^{2c}{\cal
T}(\omega ^cz_i)\tau\R) \EA\EE We thus arrive at \BE {\cal T}^{[N]}_i
=-{Q^2\over 2}{\partial^2\over \partial z_i^2} -\sum_{j\neq i}{1\over
z^N_j-z^N_i} \L( z_j z^{N-2}_i {\partial \over \partial z_j}-z^{N-1}_i
{\partial \over \partial z_i} \R)\label{(3.16)}\EE Recall that
$z^N\!\equiv\!\zeta$ can be viewed as a spectral parameter of the
$N$-KdV hierarchy, as the $N$-KdV Lax operator $L$ (see \req{QN})
satisfies $L\psi(t,z)=z^N\psi(t,z)$. In terms of these variables, the
operator \req{(3.16)} becomes, up to an overall factor, \BE -{N\over 2}
\zeta_i{\partial^2 \over \partial \zeta_i^2 }-{(N-1)\over 2}{\partial
\over \partial \zeta_i}+\sum_{j\neq i}{1\over \zeta_j -\zeta_i}
\L(\zeta_j {\partial \over \partial \zeta_j}-\zeta_i {\partial \over
\partial \zeta_i}\R)\label{(3.17)}\EE (we have set $Q^2=1$). When
imposing \V\ \cs\ on the $N$-reduced hierarchy, it is these
$\zeta_i$ that are candidates for eigenvalues of the ``source" matrix in
a \K-type matrix integral.

\section{An outlook}
\subsection{Generalized master equations from scaled \K\ matrix
integrals~?}\lvm Various aspects of the conversion of \V\ constraints
into decoupling equations deserve more study from the `Liouville' point
of view. The \K-type matrix integrals whose Ward identities coincide
with the generalized master equation, may provide a discretized
definition of the Liouville theory. More precisely, consider the matrix
integral (see \cite{[K],[MS],[KMMMZ],[IZ]}) \BE{\cal F}(\Lambda)=\int
DXe^{-\tr X^3 +\tr \Lambda X}\label{Kontsevichint}\EE where $\Lambda$ is
a `source' matrix. Then, as emphasized in the papers cited above, the
Ward identity assumes the form, \BE
\L(\sum_j\dd{\Lambda_{ij}}\dd{\Lambda_{jk}} - {1\over
3}\Lambda_{ki}\R){\cal F}(\Lambda)=0\EE \let\l=\lambda Further, the
Kontsevich integral \req{Kontsevichint} does in fact depend only on the
eigenvalues $\l_i$ of $\Lambda$. We use this to evaluate the
second-order derivative and then restrict to a diagonal $\Lambda$. This
results is \cite{[MMM]} \BE\L(\ddsc{\l_i} +\sum_{j\neq i}{1\over \l_i-
\l_j}\L(\dd{\l_i}-\dd{\l_j}\R)-{1\over 3}\l_i\R){\cal F}\{\l\}=0
\label{masterlambda}\EE

Comparing this to the master equation ${\cal T}(z_i)\tau=0$ (see
\req{Tsym}), one notices in \req{masterlambda} a puzzling term linear
in $\l$. The presence of this term, clearly, implies that $\l_j$ are
dimensionless, and therefore so would be the time parameters constructed
out of the $\l_j$ according to the \M\ formula. This is in contrast with
the fact that the KP times are naturally assigned dimensions
$t_r\sim({\rm lengh})^r$, which implies $z\sim({\rm lengh})^{-1}$.

A useful analogy is provided by the relation between the
(\V-constrained) Toda and KP hierarchies; the former is a `discrete'
\h\ with dimensionless times $x_r$, while the dimensionful KP
times follow via a scaling ansatz \cite{[S29]} \BE x_r={1\over
r}\sum_{q\geq r}{q\choose r}(-1)^{q+r}(q+1){t_{q+1}\over
\epsilon^{q+1}},\quad r\geq 1.\label{scalingtimes}\EE Taking $\epsilon$
of dimension of lengh then endows the $t_q$ with the desired dimensions.
{\it Scaling} implies taking $\epsilon\Rarrow 0$; however, the above
ansatz \req{scalingtimes} is then very singular, as it contains
arbitrarily large negative powers of $\epsilon$. One might therefore
imagine the series \req{scalingtimes} defined first for sufficiently
large $\epsilon$, and then continued to $\epsilon\Rarrow 0$. What we
will need of this scaling ansatz, is the form it takes for the
\M-transformed variables: defining\footnote{the power $-r-1$, instead
of $-r$, is in the Toda case due to the presence of the `discrete time'
$s$, which also does scale along with the $x_r$ and gives rise to the
first (the lowest) time of the KP hierarchy.} \BE x_r={1\over r}\sum_j
n_j\l_j^{-r-1}\label{Miwatransformx}\EE we see that the equation
\req{Miwatransform} for the \M\ transform of the KP times will be
recovered provided \BE\l_j=1+\epsilon z_j\label{scalinglambda}\EE and
$\l^{-r-1}_j$ are expanded in {\it negative} powers of $\epsilon$ (i.e.,
formally, for $\epsilon\Rarrow\infty$, as noted above), with the result
\BE x_r={1\over r}\sum_j n_j\sum_{q\geq r}(-1)^{q+r}{q\choose
r}(\epsilon z_j)^{-q-1}\EE

 From the above digression into the scaling limit of the Toda
hierarchy, we borrow the expression \req{scalinglambda} for the
$\l_j$~\footnote{Note, however, that the idea that the \K\ integral
\req{Kontsevichint} has a direct relevance to the {\it Toda} hierarchy,
is supported by the observation that one can introduce the discrete time
$s$ into it, simply by inserting $X^s$ as a pre-exponential
factor into the integrand.}. Substituting it into \req{masterlambda}, we
see that the linear term does not survive in the $\epsilon\Rarrow 0$
limit, while the other terms behave nicely and scale into the operator
\req{Tsym} for $\a^2=2$, \BE \ddsc{z_i}+\sum_{j\neq i}{1\over
z_j-z_i}\L(\dd{z_j}-\dd{z_i}\R)\label{simplest}\EE

Still, having $\a^2=2$ is very restrictive, and one would like to relax
this condition. It seems very encouraging in this respect that the
desired most general master operator differs from the simplest one,
\req{simplest}, by introducing {\it integer} coefficients in front of
its various terms. That is, for a $(p\pr, p)$ minimal model, the
correlation functions of a product of the dressed (as in Sect.~5)
primary fields $\Phi_{m\pr_j m_j}(z_j)$, \BE\new\BA{l}\Phi_{m\pr_j
m_j}=e^{i\alpha_{m\pr_j m_j}\varphi},\quad 1\leq m\leq p-1,\quad 1\leq
m\pr\leq p\pr -1\\ \alpha_{m\pr m}={1-m\over 2}\sqrt{2p\pr\over p}-
{1-m\pr\over 2}\sqrt{2p\over p\pr}\EA\EE
are annihilated by the master
operator \BE\new\BA{rcl}{2\over \a^2}\ddsc{z_i}&+&\sum_{j\neq
i}{1\over z_j-z_i}\L(\dd{z_j}-{\a_{m\pr_j m_j}\over\a_{21}}\dd{z_i}\R)\\
{}&=&{1\over p}\L\{p\pr\ddsc{z_i}+\sum_{j\neq i}{1\over
z_j-z_i}\L(p\dd{z_j} +
\L[m_jp\pr-m\pr_jp+p-p\pr\R]\dd{z_i}\R)\R\}\EA\label{mastermm}\EE
(of course, the insertion at $z_i$ is fixed to be $\Phi_{21}$).
Remarkably, the operator inside the curly brackets contains only integer
coefficients! It remains to be shown whether by arranging the {\it
multiplicities\/} of the $\Lambda$ eigenvalues, one can match the
coefficients in \req{mastermm}.

\subsection{W-\cs\ and higher decoupling equations}\lvm If the matter
central charge $d$ is fixed to the minimal-models series, then factoring
out the null-vector leads to a bona fide minimal model. Now, thinking in
terms of the minimal models, how can one `unkontsevich' the {\it higher}
null-vector decoupling equations~? A non-trivial realization of the
relation, noted in a somewhat different context in \cite{[BDfIZ]}, between
null-vector and W-algebra structures, seems to emerge in the present
approach as well. Recall that the symmetries of the KP hierarchy are the
implemented most easily on the dressing
operators by (see \cite{[S29]}) \BE\delta K=\L(Ke^{\varepsilon
P}\delta(v,D+\varepsilon J)K^{-1}\R)_-K\label{bilocalK}\EE (with $P$
defined in eq.\req{frLn}). This can be rewritten in a form which stresses
the `bilocal' structure, \BE\delta K=\j(t,v+(1-J)\varepsilon)\circ
D^{-1}\circ\j^\ast(t,v-J\varepsilon)K\label{bilocalpsi}\EE whence it is
immediate to derive the corresponding variation of the tau function,
\BE\new\BA{rcl}\delta\tau&=&{1\over\varepsilon}\exp\biggl\{\xi(t,v+(1-J)
\varepsilon) -\xi(t,v-J\varepsilon)\biggr\}\\{}&\times&\exp\L\{\sum_{k\geq
1}{\varepsilon^k\over k}\L(J^k-(J-1)^k\R)\sum_{r\geq
1}v^{-r-k}{k+r-1\choose r} \dd{t_r}\R\}\tau\EA\label{bilocaltau}\EE Now,
the set of the higher \cs\ implied by the \V\ \cs\ reads \BE{1\over 2\pi
i}\oint{dv\over v-z}\L(K\L(e^{\varepsilon P}\delta(v,D+\varepsilon J) -
\delta(v,D)\R)K^{-1}\R)_-=0.\EE Repeating the steps \req{bilocalpsi} and
\req{bilocaltau}, and performing the substitution \req{Miwatransform} in
the $\xi$-factor, we arrive at, \BE\new\BA{rcl}{\frac{1}{2\pi
i}\oint{dv\over v-z_i}{1\over\varepsilon}\L(\exp\L\{-\sum_{k\geq
1}{\varepsilon^k\over k}\L(J^k-(J-1)^k\R)\sum_j
{n_j\over(v-z_j)^k}\R\}\R.}&{}&{}\\ \L.\times\exp\L\{\sum_{k\geq
1}{\varepsilon^k\over k}\L(J^k-(J-1)^k\R)\sum_{r\geq
1}v^{-r-k}{k+r-1\choose r} \dd{t_r}\R\}-1\R)\tau&=&\sum_{r\geq
1}z_i^{-r-1}\dd{t_r}\tau\EA\label{conj}\EE The conjecture is that by
expanding an equation of the type of \req{conj} in powers of $\varepsilon$
(this equation by itself seems too naive to be {\it the} one needed), and
for certain $J$-dependent values of $n_i$, one would arrive at a set of
the higher decoupling equations. From these, expressions for the null
vectors could in turn be extracted. Besides the level-2/\V\ case
considered in this paper, a version of the level-3 decoupling equation was
shown recently \cite{[GS]} to correspond via the \K-\M\ transform to
$W^{(3)}$ \cs\ on the KP \h.

\subsection{Master equations on \RS s}\lvm
\def\T{\hat\tau} A comment is in order concerning the `global' structure
(or, the `boundary conditions') of the \K-transformed tau functions. The
master operator and hence eq.\req{decouplingn}, with the characteristic
$(z_i-z_j)$ de\-no\-mi\-na\-tors, are written down in a given coordinate
system. The coordinate patch must cover the neighbourhood of the `infinity'
which contains all the points $z_j$. Otherwise, it may be an arbitrary
neighbourhood on a \RS. That is, `closing up' the neighbourhood to the
Rimann sphere ${\open CP}^1$, we imply certain boundary conditions on
the tau function subjected to the master equation. If, however,
the equation can be {\it covariantly\/} carried over to the whole of a
\RS\ glued to the patch, then it should be possible to impose the
corresponding boundary conditions on $\tau$, i.e, consider it as a
solution on the \RS.

Indeed, let $S$ be a \RS\ of genus $g$ and $E(P,Q)$ its associated prime
form \cite{[F]}. For $P$ and $Q$ both in the coordinate neighbourhood,
one has \BE\ln E(z,y)=\ln(z-y)+\sum_{m,n\geq 1}Q_{mn}{z^{-m}y^{-n}\over
mn},\EE (the coordinate system is centered at an $R\in S$,
$z^{-1}(R)=0$.) Choose $(g-1)Q\equiv(g-1)(2J-1)$
points $P_\a\in S$. Let $\tau(t)$
denote, as before, a KP tau function constrained with the help of the
\V\ generators ${\Sc L}_{\geq-1}$, eq.\req{Lontau}. To separate the
factors that carry a dependence on the details (such as the center etc.)
of the coordinate neighbourhood, define a function $\T$ by (cf.
\cite{[S12]}), \BE\new\BA{rcl}\tau(t)&=&{\cal N}e^{\half{\sum_{r,s,\geq
1}Q_{rs}t_rt_s}} \prod_{\a=1}^{(g-1)Q}e^{-\sum_{r\geq 1}rt_r
\int_z^{P_\a}\o^{(r)}}\\ {}&\times&\exp Q\L\{{1\over 2\pi i}\sum_{r\geq
1}rt_r\sum_{i=1}^g\oint_{a_i}\o_i(u)\int\limits_z^u\o^{(r)} +
\sum_{r\geq 1}t_r\L(-z^r+\sum_{s\geq 1}Q_{rs}{z^{-s}\over s}\R)\R\}\cdot
\T\EA\label{tauT}\EE where $\o^{(r)}$ are the meromorphic differentials
with a pole at $R$, $\o^{(r)}(z)\equiv\o^{(r)}_R(z)$ where, more
generally, \BE\o^{(r)}_a(z)={1\over r!}{\rm d}_z{\d^r\over \d a^r}\ln
E(a,z)\EE is the meromorphic differential with a pole at $a$ of order
$r+1$ and holomorphic everywhere else on $S$. Further, $\o_i$,
$i=1,\ldots,g$ are the holomorphic differentials normalized by
$\oint_{a_i}\o_i=0$ where $a_i$ are the $a$-cycles in homology. The RHS
of \req{tauT} is independent of a point with coordinate $z$. Finally,
${\cal N}$ is a normalization factor, \BE{\cal
N}=\prod_{\a<\beta}E(P_\a,P_\beta)\prod_\beta\sigma(P_\beta)^Q,\EE where
\BE\sigma(u)=\exp{-1\over 2\pi i}\sum_{i=1}^g\oint_{a_i}\o_i(y)\ln
E(y,u)\EE is Fay's $g/2$-differential.

Performing in \req{tauT} the substitution \req{Miwatransform}, we bring
it to the form, \BE\tau\{z_j\}=\prod_{j<k}\L({E(z_j,z_k)\over
z_j-z_k}\R)^{n_jn_k}\prod_{\a=1}^{(g-1)Q}\prod_j
{E(R,P_\a)^{n_j}\sigma(R)^{QN_j}\over
E(z_j,P_\a)^{n_j}\sigma(z_j)^{Qn_j}}
\prod_{\a<\beta}E(P_\a,P_\beta)\prod_\beta\sigma(P_\beta)^Q\cdot\T\{z_j\}\EE
The dependence on the reference point $R$ drops out for the `neutral'
sets $\sum_jn_j=0$. Then the points $P_\a$ can be viewed as an addition
to the $\{z_j\}$, entering with prescribed Miwa coefficients \BE
n_\a=-1,\quad \a=1,\ldots,(g-1)(2J-1).\EE

Thus, the \M\ transform associated to a given \RS\ and a coordinate
neighbourhood chosen on it, reads \BE
t_r={1\over r}\sum_j n_jz^{-r}_j,\quad \sum_jn_j=-(g-1)(2J-1).\EE The
number $(g-1)(2J-1)$ is of course the RHS of the Riemann-Roch theorem
(and should therefore be modified accordingly for $g=0,~1$ and $Q=1$).

In the master equation which holds for $\T\{z_j\}$ all the `bare'
$(z_i-z_j)^{-1}$ get replaced by the corresponding meromorphic
differentials constructed out of $E(z_i,z_j)$ (other terms appear as
well, among them the projective connection on $S$ \cite{[F]}). This
equation being in this sense {\it covariant,\/} means that $\T\{z_j\}$
may be extended to the rest of $S$.

We have thus shown that the master equation can be extended onto
higher-genus \RS s. A field-theoretic construction providing a solution
on such a surface is still to be clarified. As to the {\it equation\/}
alone, we have seen that, as expected, ``\RS s of different genera solve
equally well the integrable equations''.

\bigskip
{\bf {\sc Acknowledgements}}. I would like to thank S.~Kharchev,
A.~Mironov and A.~Zabrodin for interesting conversations. I am grateful
to A.~Subbotin for useful remarks on the manuscript.

       \end{document}